\documentclass[aps,prl,reprint,groupedaddress,showpacs,raggedbottom]{revtex4-1}

\usepackage{amsmath}
\usepackage{amssymb}
\usepackage{graphicx}
\usepackage{verbatim}
\usepackage{lineno}

\pdfoutput=1

\newcommand{\eq}{\begin{equation}}
\newcommand{\eeq}{\end{equation}}
\newcommand{\ket}[1]{\left|#1\right\rangle}
\newcommand{\bra}[1]{\left\langle #1\right|}

\begin{document}

\title{Entanglement of distinguishable quantum memories}

\author{G. Vittorini}
\email{vittorgd@umd.edu}
\author{D. Hucul}
\author{I.V. Inlek}
\author{C. Crocker}
\author{C. Monroe}
\affiliation{Joint Quantum Institute, University of Maryland Department of Physics and National Institute of Standards and Technology, College Park, Maryland 20742, USA}

\date{xxxxxxx}

\begin{abstract} 

Time-resolved photon detection can be used to generate entanglement between distinguishable photons. This technique can be extended to entangle quantum memories that emit photons with different frequencies and identical temporal profiles without the loss of entanglement rate or fidelity. We experimentally realize this process using remotely trapped $^{171}$Yb$^+$ ions where heralded entanglement is generated by interfering distinguishable photons. This technique may be necessary for future modular quantum systems and networks that are composed of heterogeneous qubits.

\end{abstract}

\maketitle


The entanglement of remote quantum memories via photons is an enabling technology for modular quantum computing, transmission of quantum information, and networked quantum timekeeping~\cite{Cirac97,Briegel98,Duan01,Komar13,Monroe14}. All of these applications rely on the excellent control and long-lived coherence properties of quantum memories and the ease with which photons can be used to distribute both entanglement and quantum information. 

Currently, photon-mediated entanglement of remote quantum memories has only been achieved using heralded schemes; atomic ensembles~\cite{Chou05}, trapped ions~\cite{Moehring07}, single neutral atoms in optical cavities~\cite{Noelleke13}, and nitrogen vacancy centers in diamond~\cite{Bernien13} have all been entangled in this manner. However, heralded entanglement is possible in these systems because the memories inherently emit indistinguishable photons or can be manipulated to do so. This requirement of photon indistinguishability is a major impediment to the construction of heterogeneous quantum networks and also hinders the entanglement of similar memories whose emission frequencies differ due to variations in local environment or fabrication. Recently, photons with frequencies that differ by many linewidths have been entangled using time-resolved detection and active feedforward~\cite{Zhao14}. Here, we extend this technique to entangle distinguishable remote quantum memories by interference of distinguishable photons. 

In order to generate heralded entanglement of remote quantum memories, photons emitted from each memory enter a partial Bell state analyzer where they interfere on a beamsplitter and are subsequently detected, projecting the memories into a corresponding entangled state. If the photons are identical, the memories will be projected into a known Bell state~\cite{Simon03}. However, if the photons are distinguishable, the resulting entangled memory state will have some additional phase factor or unequal probability amplitudes. Here we assume that quantum memories labeled A and B emit photons with identical temporal profiles but frequencies that differ by $\Delta \omega$ for a given polarization~\cite{Matsukevich08}. These two photons are detected within the Bell state analyzer a time $\Delta t$ apart, projecting the memories into a Bell state with a time-dependent phase, e.g.,  ${\frac{1}{\sqrt{2}}(\ket{01}+e^{-i\Delta\omega\Delta t}\ket{10})}$, where $\{\ket{0}$,$\ket{1}\}$ are the computational basis states of each memory.  The temporal resolution $t_r$ of the photon detection circuit determines how precisely the phase $\Delta \omega \Delta t$ is known since the phase is probabilistically distributed over an interval of characteristic width $\Delta \omega t_r$. Thus, if $t_r \gtrsim 2 \pi/\Delta \omega$, then the ambiguity of the phase results in decoherence of the entangled memory state.

One solution is to use time-resolved photon detection~\cite{Legero03, Metz08} where $t_r \ll 2\pi /\Delta \omega$. In this limit, the phase of the memory state $\Delta \omega t_r \ll 2\pi$ and therefore well-defined for a single heralded entanglement event. However, averaging over all heralded entanglement events will lead to decoherence due to the probabilistic distribution of photon detection times $\Delta t$. This decoherence can be reduced by postselecting entanglement events in which $\Delta t \ll 2\pi / \Delta \omega$ at the cost of entanglement rate~\cite{Dyckovsky12}. Alternatively, if $\Delta \omega$ is known, $\Delta t$ can be fed forward to subsequent operations to maintain high-fidelity, constant-phase memory entanglement without sacrificing rate~\cite{Zhao14}.

In this Rapid Communication, we entangle two remote quantum memories that emit distinguishable photons by utilizing time-resolved detection of photonic polarization qubits. We reveal the time-dependent nature of the resulting entangled memory state and observe that temporally filtering detection events results in maximized state fidelity. Finally, we show how feedforward can be used to generate high fidelity entanglement without a reduction in entanglement rate.

\begin{figure}
\includegraphics[width = 3.375 in]{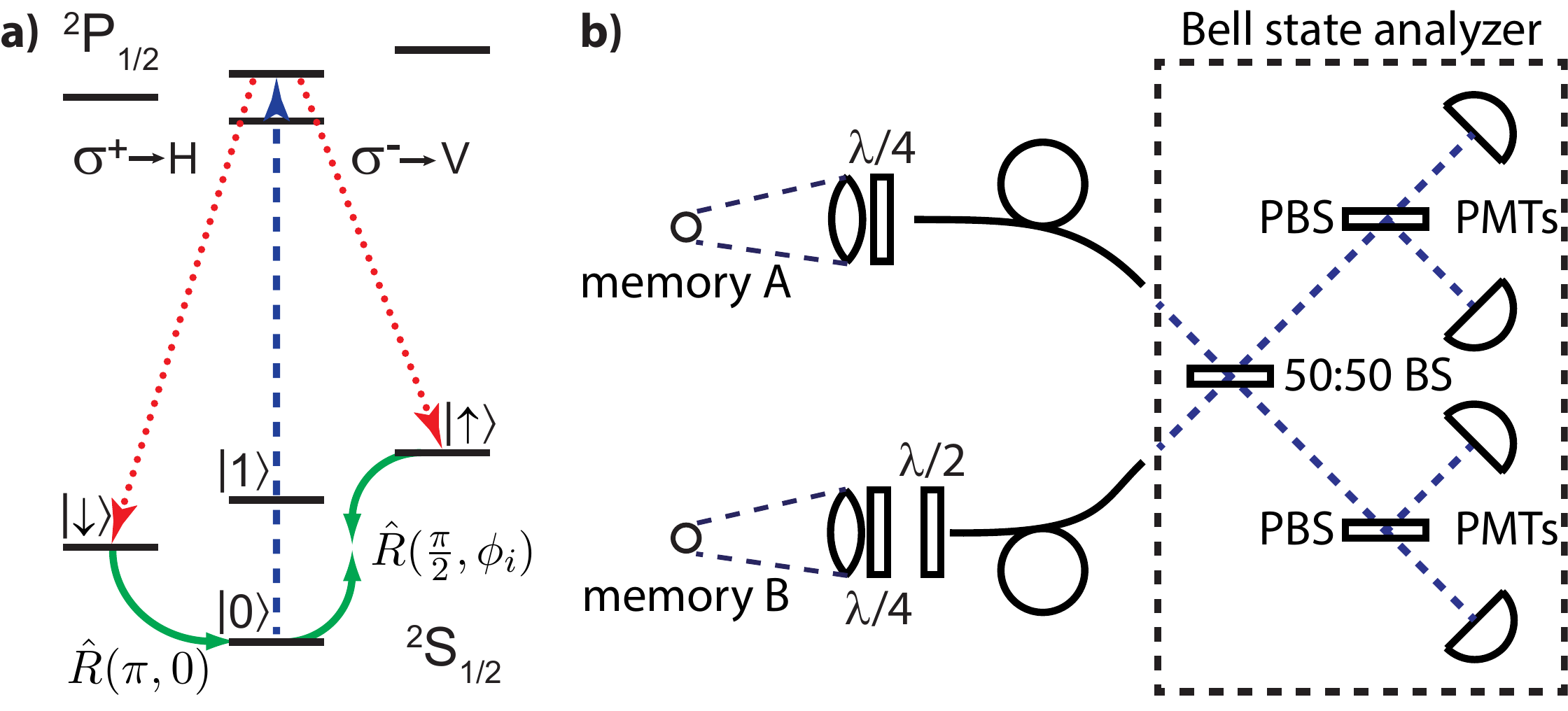}
\caption{(Color online). (a) The energy level diagram of the ${}^{171}$Yb$^{+}$ quantum memory. Photon generation~(blue dashed) and collection~(red dotted) transitions are labeled along with the microwave transitions~(green solid) used to analyze the entangled memory state. $\sigma^\pm$ polarized light is emitted by each memory before being converted to the linear basis by a $\lambda/4$ waveplate. (b) Experimental setup including quantum memories, photon collection optics, and partial Bell state analyzer. The analyzer consists of a 50:50 non-polarizing beamsplitter (50:50~BS), two polarizing beamsplitters (PBS) and four photomultiplier tubes (PMT).}
\label{fig:exp_system}
\end{figure}

We use trapped ${}^{171}$Yb$^{+}$ atoms as quantum memories [Fig.~\ref{fig:exp_system}a]. A magnetic field at each memory provides a quantization axis while standard photon scattering methods are used for Doppler cooling, state initialization and detection~\cite{Olmschenk07}. We apply microwave fields to manipulate the state of each atom within the ${}^2S_{1/2}$ ground state manifold.

The quantum memories are entangled using photonic polarization qubits~\cite{Matsukevich08}. To generate a single photon from an atom, it is initially excited using a $\pi$ polarized, 2~ps pulse from a Ti:sapphire laser resonant with the ${}^2S_{1/2} \rightarrow {}^2P_{1/2}$ transition [Fig.~\ref{fig:exp_system}a]. The memory then emits a single photon with a $\tau = 8.1$~ns decaying exponential temporal profile. This photon's polarization is entangled with its parent atom's state due to atomic selection rules~\cite{Matsukevich08}. The photon is then collected along the quantization axis by a numerical aperture NA~$ = 0.65$ objective, coupled to a single-mode fiber mounted in a manual fiber polarization controller, and directed to a partial Bell state analyzer for detection [Fig.~\ref{fig:exp_system}b]. Though photons of any polarization are emitted, $\pi$ polarized photons do not couple to the fiber due to the electric field's rotational cylindrical symmetry~\cite{Kim11}. A $\lambda/4$ waveplate just prior to the fiber rotates the $\sigma^\pm$ polarized photons to the linear basis~$\{\ket{H},\ket{V}\}$ and the fiber is strained to maintain the photon polarization. The resulting memory-photon entangled state after the fiber is $\frac{1}{\sqrt{2}}(\ket{\uparrow}\ket{V}-i\ket{\downarrow}\ket{H})$, where $\{\ket{\uparrow},\ket{\downarrow}\}$ denote the ${}^2S_{1/2}\ket{F=1,m_F=\pm 1}$ atomic states (see Fig.~\ref{fig:exp_system}a).

Within the Bell state analyzer, photons from each memory coincidentally impinge upon the beamsplitter before being sorted in the linear basis~$\{\ket{H},\ket{V}\}$ by thin film polarizers and detected with a time difference $\Delta t$ during a coincidence window $T = 60$~ns. The start of $T$ is triggered by the ultrafast excitation pulse and the window length is set to encompass ${>}99\%$ of the photonic temporal profile.  Clicks from specific detector pairs indicate that the joint photon state prior to detecting a photon is ideally a Bell state with a time-dependent phase. This detection event projects the quantum memories into the corresponding entangled state
\begin{equation}
\ket{\Psi} = \frac{1}{\sqrt{2}}(\ket{\downarrow\uparrow}+\textrm{e}^{-i (\Delta\omega \Delta t+2\Delta\omega t'-\phi_D+\phi_0)}\ket{\uparrow\downarrow})\\
\end{equation}
where $t'$ is the time elapsed following the detection of the second photon, $\phi_D$ is $0$ or $\pi$ depending on which pair of detectors registers the photons~\cite{Simon03} and $\phi_0$ is a stable intermemory phase~\cite{Hucul14}. The probability of two photon collection and detection during a window $T = 60$~ns is $\sim\!10^{-5}$ resulting in an entanglement rate of order Hz~\cite{Hucul14}.

\begin{figure}
\includegraphics[width = 3.375 in]{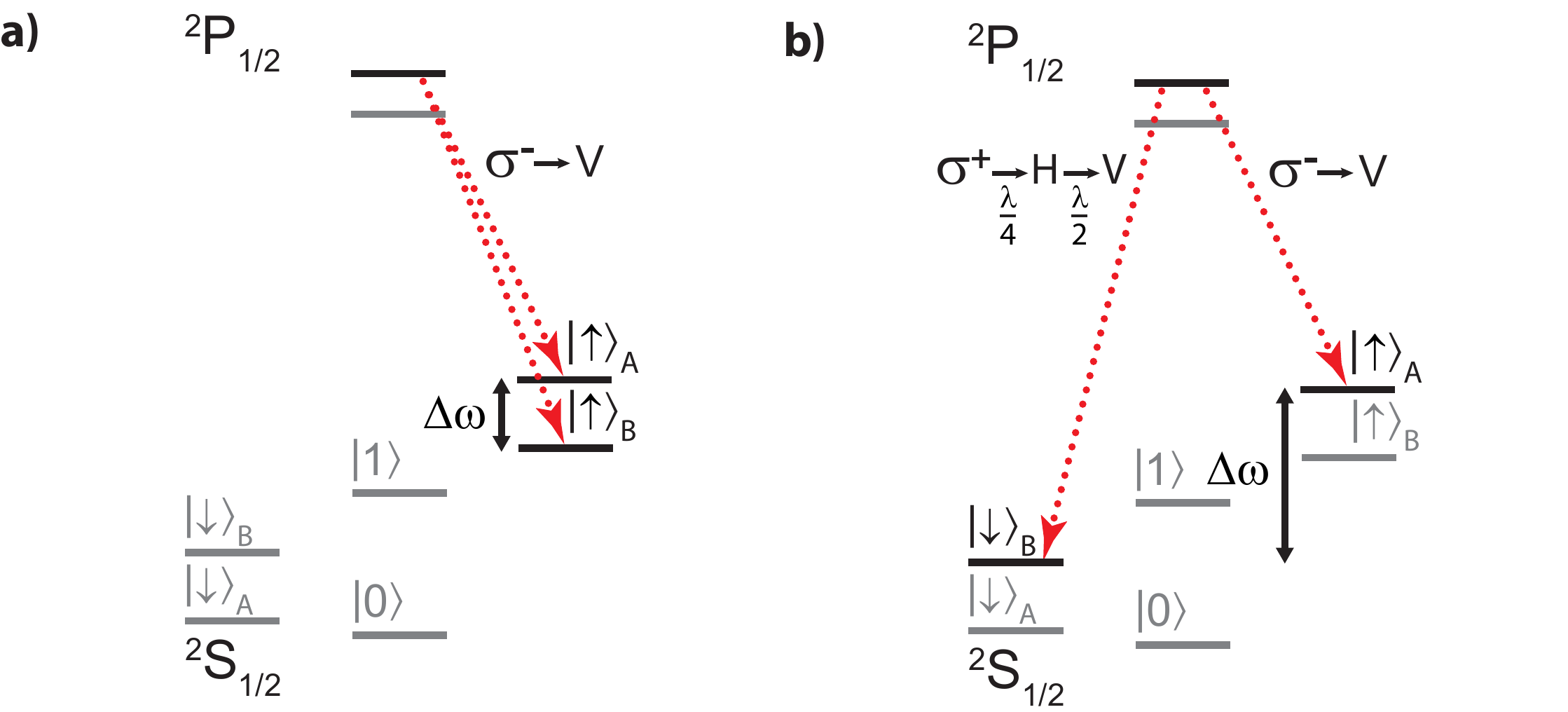}
\caption{(Color online). (a) In the conventional memory-memory entanglement configuration, the polarization of photons from both memories are associated with the same atomic state. The resulting photon frequency difference $\Delta \omega$ is limited by the largest magnetic field difference we can apply across the two memories. The ket subscripts represent which memory the state is associated with. (b) By introducing a $\lambda/2$ waveplate to one photon path and rotating that photon's polarization to the orthogonal linear state, the polarizations of the two photons are now associated with atomic states with opposite Zeeman shifts. This technique maximizes $\Delta \omega$ for a given maximum magnetic field magnitude.}
\label{fig:pol_map}
\end{figure}

For our conventional memory-memory entanglement configuration, we apply different magnetic field magnitudes at each atom to adjust the photon frequency difference $\Delta \omega$ [Fig.~\ref{fig:pol_map}a]. However, a minimum magnetic field must be applied to eliminate coherent dark states which reduce Doppler cooling efficiency~\cite{Berkeland02}. Large maximum fields can be generated in principle but shifts of the magnetic field sensitive levels on the order of the transition linewidth complicate cooling, state preparation, and state detection.  An alternative approach makes use of similar magnetic field magnitudes at each quantum memory and a $\lambda/2$ waveplate in memory B's photon path [Fig.~\ref{fig:exp_system}b]. In this configuration, the waveplate rotates the photon polarization for a single memory to the orthogonal linear state to map $\frac{1}{\sqrt{2}}(\ket{\uparrow}\ket{V}-i\ket{\downarrow}\ket{H}) \rightarrow \frac{1}{\sqrt{2}}(\ket{\uparrow}\ket{H}-i\ket{\downarrow}\ket{V})$. Photons with identical polarization from different atoms now have a frequency difference $\Delta \omega$ equal to the Zeeman splitting between the $\ket{\uparrow}_\textrm{A}$ and $\ket{\downarrow}_\textrm{B}$ states for a given magnetic field magnitude [Fig.~\ref{fig:pol_map}b]. Upon detection of both photons by an appropriate detector pair, the ions are projected into the entangled state
\begin{equation}
\ket{\Phi} = \ket{\downarrow\downarrow}-\textrm{e}^{i (\Delta\omega \Delta t+2\Delta \omega t'+\phi_D+\phi_0)}\ket{\uparrow\uparrow}.
\end{equation}

In order to analyze the entangled memory state, we experimentally extract elements from the density matrix by applying rotations concurrently to both memories [Fig.~\ref{fig:exp_system}a]. Since fluorescence detection does not distinguish between the $\ket{\downarrow}$ and $\ket{\uparrow}$ states, a microwave $\pi$-pulse first transfers any atomic population from $\ket{\downarrow}$ to $\ket{0}$ in both ion traps where $\ket{0}$ denotes the ${}^2S_{1/2}\ket{F=0,m_F=0}$ atomic ground state. We then use state-dependent fluorescence to measure the populations $P_{\ket{00}}$ , $P_{\ket{0\uparrow}}$, $P_{\ket{\uparrow0}}$, and $P_{\ket{\uparrow\uparrow}}$. To determine the coherences $\rho_{\ket{\Psi}}$ and $\rho_{\ket{\Phi}}$, an analysis pulse on each memory follows the initial transfer pulse~\cite{Sackett00}. The analysis pulse consists of a rotation $\hat{R}(\frac{\pi}{2},\phi_a)$ ($\hat{R}(\frac{\pi}{2},\phi_b)$) resonant with the $\ket{0}\rightarrow\ket{\uparrow}$ transition of memory A (B) where $\phi_i$ is the phase of the microwave radiation applied to memory $i$. We measure the probability of being in an odd parity state $P^o = P_{\ket{0\uparrow}}+P_{\ket{\uparrow0}}$ for a variety of analysis phases $\phi_a$, $\phi_b$ using state-dependent fluorescence. To extract $\Pi_{\ket{i}} = |\rho_{\ket{i}}|$, the data are fit to 
\begin{equation}
\begin{split}
P^o_{\ket{\Psi}} &= \frac{1}{2} - \Pi_{\ket{\Psi}} \cos( \phi_a - \phi_b + \Delta \omega \Delta t -\phi_D + \phi_0)\\
P^o_{\ket{\Phi}} &= \frac{1}{2} - \Pi_{\ket{\Phi}}\cos(\phi_a + \phi_b -\Delta \omega \Delta t -\phi_D - \phi_0).
\label{eq:odd_parity}
\end{split}
\end{equation}
By choosing $\phi_a = -\phi_b$ we measure $P^o_{\ket{\Psi}}$ exclusively and with $\phi_a = \phi_b$ we measure $P^o_{\ket{\Phi}}$. We then calculate the fidelity $\bra{\Psi}\hat{\rho}\ket{\Psi} = \frac{1}{2}(P_{\ket{0\uparrow}}+P_{\ket{\uparrow0}})+\Pi_{\ket{\Psi}}$ or $\bra{\Phi}\hat{\rho}\ket{\Phi} = \frac{1}{2}(P_{\ket{0\uparrow}}+P_{\ket{\uparrow0}})+\Pi_{\ket{\Phi}}$ to verify the creation of the desired entangled state.

\begin{figure}
\includegraphics[width = 3.375 in]{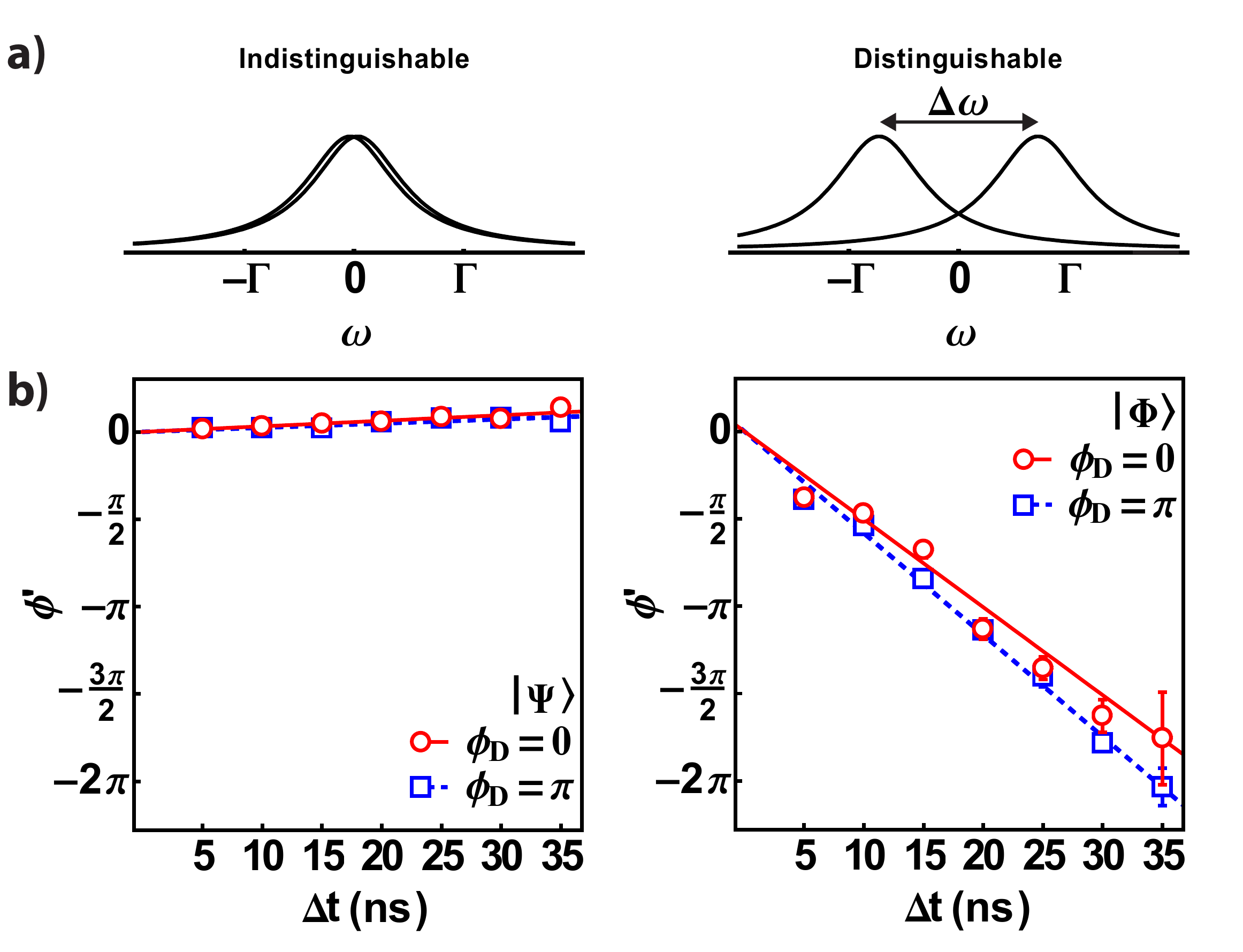}
\caption{(Color online). (a) Frequency lineshapes for indistinguishable ($\Delta \omega = 2 \pi \times 1.35$~MHz) and distinguishable ($\Delta \omega = 2 \pi \times 28.35$~MHz) photons with linewidth $\Gamma = 1/\tau = 2\pi \times 19.6$~MHz. (b) Extracted time-dependent phase evolution of $P^o$ oscillations for the $\ket{\Psi}$ and $\ket{\Phi}$ memory states. We find $\Delta \omega = 2 \pi \times 1.4(2)$~MHz for $\ket{\Psi}$ and $\Delta \omega = 2 \pi \times 27.1(1.7)$~MHz for $\ket{\Phi}$. The ${\phi' = \pm \Delta \omega \Delta t + \phi_0}$ data have been offset such that $\phi_0=0$ and the error bars correspond to the standard error of the mean.}
\label{fig:time_evo}
\end{figure}


We analyze the evolution of the entangled memory state for indistinguishable and distinguishable photons using a photon detection circuit with temporal resolution $t_r = 5$~ns. With similar magnetic field amplitudes at each memory, we set the angle between the slow axis of the $\lambda/2$ waveplate and the $V$ polarization direction to be $0$ ($\pi/4$) to generate indistinguishable (distinguishable) photons with $\Delta \omega = 2\pi \times 1.35$~MHz~$= 0.069/\tau$ ($\Delta \omega = 2\pi \times 28.35$~MHz~$= 1.45/\tau$) [Fig.~\ref{fig:time_evo}a]. We separate the $P^o$ data according to the $\Delta t$ recorded by our photon detection circuit and extract the phase $\phi' = \pm\Delta \omega \Delta t + \phi_0$ by fitting the data to Eq.~\ref{eq:odd_parity}. We calculate $\Delta \omega$ from the slope of $\phi '$ versus $\Delta t$ and average the $\Delta \omega$ values for $\phi_D = 0,\pi$. For the indistinguishable (distinguishable) case, we find a mean experimental phase evolution of $\Delta \omega = 2\pi \times 1.4(2)$~MHz ($\Delta \omega = 2\pi \times 27.1(1.7)$~MHz) which agrees with the measured qubit splitting [Fig.~\ref{fig:time_evo}b]. Furthermore, the accrued phase has the expected sign.

\begin{figure}
\includegraphics[width = 3.375 in]{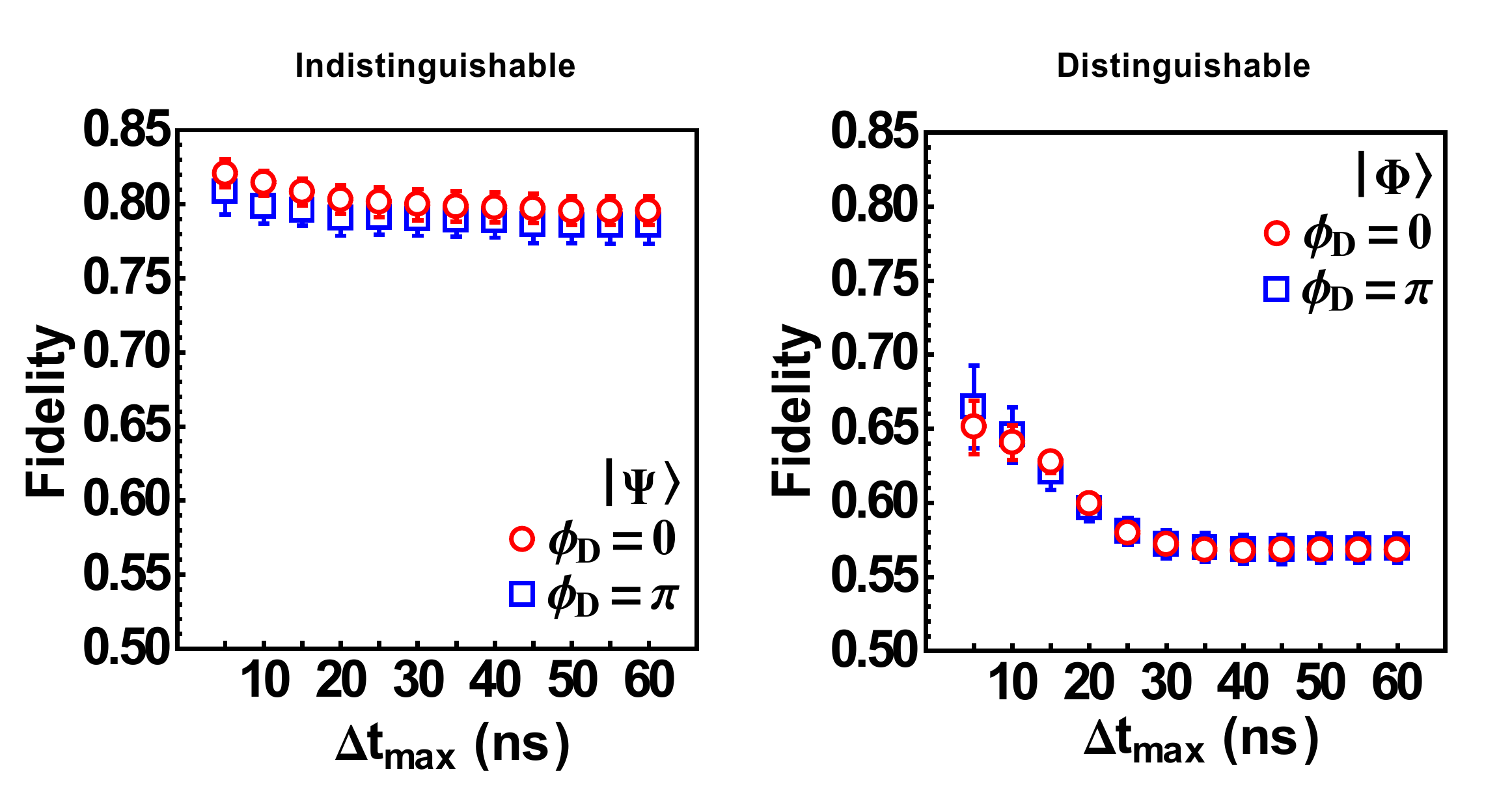}
\caption{(Color online). Entangled memory state fidelity vs. maximum time  between detector clicks during an entanglement event. For the indistinguishable case where the joint memory state is $\ket{\Psi}$, the fidelity is nearly constant for all values of $\Delta t_{\textrm{max}}$. For the $\ket{\Phi}$ state, which is associated with distinguishable photons, there is a clear increase in fidelity for $\Delta t_{\textrm{max}} < 2\pi / \Delta \omega \approx 35$~ns. Error bars correspond to the standard error of the mean.}
\label{fig:fidelity}
\end{figure}

\begin{figure*}
\includegraphics[width = 6.75 in]{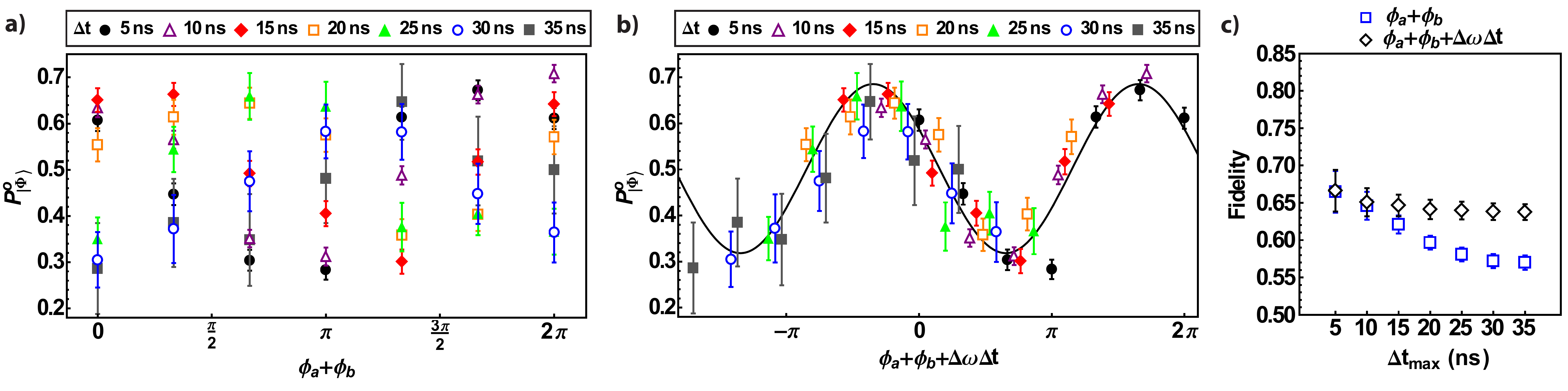}
\caption{(Color online). a) $P^o_{\ket{\Phi}}$ ($\phi_D = \pi$) as a function of analysis pulse phases $\phi_a$ and $\phi_b$ for varying values of photon detection time difference $\Delta t$. The sinusoids are incoherent due to their time-dependent phase shifts. b) We postprocess the entangled state by shifting the analysis phases $\phi_a + \phi_b \rightarrow \phi_a + \phi_b + \Delta \omega \Delta t$ using the experimentally measured photon frequency difference $\Delta \omega$ and $\Delta t$. The resulting parity oscillations are in phase. Here, the fit serves as a guide to the eye. c) Fidelity of the entangled memory state for the original (blue squares) and phase-shifted (black diamonds) data. Fidelity loss due to the time-dependent phase is significantly reduced for the latter. All error bars correspond to the standard error of the mean.}
\label{fig:phase_shift}
\end{figure*}

Averaging over the phase evolution of many entanglement events results in decoherence of the joint memory state.  To verify this behavior, we analyze the resulting state for all entanglement events in which $\Delta t \leqslant \Delta t_{\textrm{max}}$, where $\Delta t_{\textrm{max}} \leqslant T$ is a variable, maximum photon interarrival time [Fig.~\ref{fig:fidelity}]. We expect no significant phase evolution for the indistinguishable case and thus minimal loss of fidelity for any $\Delta t_{\textrm{max}}$. For the distinguishable case, the entangled state phase and thus $P^o_{\ket{\Phi}}$ advances by $2\pi$ for $\Delta t = 2\pi/\omega \approx 35$~ns. Therefore as $\Delta t_{\textrm{max}}$ increases, the entangled state fidelity decreases due to the out-of-phase contributions of entangled states with different $\Delta t$. In this experiment, the fidelity asymptotes to a value larger than the mixed state limit of $0.5$ as the majority of entanglement events occur with $\Delta t \leqslant 10$~ns due to the exponential distribution of $\Delta t$. For $\Delta t\leqslant5$~ns, the fidelity $\bra{\Phi}\hat{\rho}\ket{\Phi}$ is maximized despite the distinguishability of the emitted photons but it does not reach $\bra{\Phi}\hat{\rho}\ket{\Phi}$. The observed discrepancy may be explained by the protection against dephasing provided by the $\ket{\Psi}$ state~\cite{Kielpinski01}. Since the $\ket{\downarrow}$ and $\ket{\uparrow}$ states are first-order sensitive to magnetic field fluctuations, significant dephasing of the $\ket{\Phi}$ state occurs due to common-mode magnetic field noise during the ${\sim}50$~$\mu$s necessary for the transfer and analysis pulses. Additionally, the $\ket{\Phi}$ state is more sensitive to ns jitter in the photon detection and analysis electronics due to the larger value of $\Delta \omega$.

In order to set the phase of the entangled memory state when $t_r$ is sufficiently small, either $\Delta t_{\textrm{max}}$ must also be small or $\Delta t$ used to feedforward a phase adjustment to the memory state~\cite{Zhao14}. By restricting $\Delta t_{\textrm{max}} \ll 2 \pi / \Delta \omega$, the maximum possible entangled memory state phase evolution is $\Delta \omega \Delta t_{\textrm{max}}$. However, the number of entanglement events decreases due to the exponential distribution of $\Delta t$. Taking $\Delta t_{\textrm{max}} = t_r = 5$~ns, our experimental entanglement rate $R = 0.2 R_0$ where $R_0$ is the rate when all entanglement events are accepted. By setting $\Delta t_{\textrm{max}} = t_r$, we can maintain a large coincidence window $T$ which spans the photonic temporal profile and does not temporally filter the coincident detection events. We then only select events in which $\Delta t = 5$~ns from within $T$. This is an improvement over a simple gating procedure~\cite{Dyckovsky12} in which the coincidence window $T = t_r$ and $R = 0.07 R_0$ in our experiment.

Feeding $\Delta t$ forward to subsequent operations eliminates the need for postselection at the cost of increased overhead. In this experiment, the simplest method to convert $\Delta \omega \Delta t$ to a constant phase $\phi_c$ consists of waiting a time $t' = \frac{1}{2\Delta \omega}(\phi_c\pm\phi_D-\phi_0-\Delta \omega \Delta t)$ following an entanglement event where the top sign is associated with $\ket{\Phi}$ and the bottom with $\ket{\Psi}$~\footnote{By applying the modulo $2\pi$ operation to $(\phi_c\pm\phi_D-\phi_0-\Delta \omega \Delta t)$, one can determine the minimum positive value of $t'$ necessary to generate the desired $\phi_c$.}. Such a wait operation could be performed in $t' \leqslant 370$~ns ($t' \leqslant 18$~ns) for any $\Delta t$ in the indistinguishable (distinguishable) photon case. Alternative approaches include applying a $\hat{\sigma}_z$ interaction for an appropriate time to a single memory or recording $\Delta t$ and modifying the phase of subsequent rotations of the entangled state. For trapped ions, this feed forward operation can be completed many orders of magnitude faster than entanglement is generated~\cite{Hucul14} resulting in minimal additional overhead. However, any feedforward operation will require a temporal resolution $\ll\!\pi/\Delta \omega$ in order to track the entangled state phase $2\Delta \omega t'$ and faithfully produce $\phi_c$. We postprocess the entangled memory state $\ket{\Phi}$ for $\phi_D = \pi$ by shifting the phase of each $P^o_{\ket{\Phi}}$ oscillation by the known phase accumulation $\Delta \omega \Delta t$ such that $\phi_a + \phi_b \rightarrow \phi_a + \phi_b + \Delta \omega \Delta t$. The resulting $P^o_{\ket{\Phi}}$ curves are coherent for any photon detection time difference $\Delta t$ and the fidelity is maximized [Fig.~\ref{fig:phase_shift}].


We have utilized time-resolved photon detection to demonstrate entanglement of quantum memories that emit distinguishable photons. We filter entanglement events to reduce decoherence and generate uniform entanglement without sacrificing entanglement rate. This system is amenable to the use of feedforward to generate entangled states with no loss in entanglement rate due to postselection or postprocessing. These techniques can be applied to a heterogeneous quantum network or other modular quantum system constructed of nonidentical components without the need for the modification of individual electromagnetic or strain environments. Instead, the quality of entanglement within the system is dependent on the speed and phase noise of the detection electronics and their clock. Photon detectors with jitter of order 10 ps, coincident detection electronics with ps resolution, and stable oscillators with frequencies of order 10 GHz are currently available and could ideally be used to generate entanglement between memories with identical temporal profiles and $\Delta \omega \sim 2 \pi \times 1$~GHz via feedforward. Moreover, high fidelity entanglement of quantum memories that may differ in physical origin, such as trapped ions and quantum dots, is possible by interfering photons with arbitrary frequency difference as long as the detection circuit temporal resolution $t_r \ll 2\pi/\Delta \omega$ and the photon temporal profiles are sufficiently matched~\footnote{If the emitted photon temporal profiles differ, then photon arrival information may be used to distinguish the two photons. This temporal profile mismatch modifies the probability amplitudes of the memory state, not the phase, and can result in the creation of a completely unentangled state when the lifetimes differ by a large degree i.e., $\tau_a \ll \tau_b$. In this case, one solution is to set $\Delta t_{\textrm{max}} \ll \tau_a$ to maximize the fidelity of the entangled memory state at the expense of entanglement rate.}. As long as the detection circuit temporal resolution meets this criteria, then all phase evolution of the entangled state can be tracked and corrected to generate a coherent ensemble of entangled memory states.

\begin{acknowledgments}
We thank K. Wright and J.D. Wong-Campos for critical review of the manuscript. This work was supported by the Intelligence Advanced Research Projects Activity, the Army Research Office MURI Program on Hybrid Quantum Optical Circuits, Defense Advanced Research Projects Agency SPARQC and the NSF Physics Frontier Center at JQI.
\end{acknowledgments}

\bibliography{NID_emitters}

\end{document}